\def\.#1{\mathaccent 95#1}
\def\^#1{\mathaccent 94 #1}
\def\~#1{\mathaccent "7E #1}

\def\unit{{\cal{I}}}

  \def\ket{\vert \vert  \{ \emptyset \} \rangle}
  \def\ket2{\vert \vert \otimes \{ R \} \rangle}

\def\.#1{\mathaccent 95#1}
\def\^#1{\mathaccent 94 #1}
\def\~#1{\mathaccent "7E #1}

\def\eq{\enskip =\enskip}
\def\pls{\enskip +\enskip}
\def\mns{\enskip -\enskip}

\def\unit{{\cal I}}

  \def\ket{\vert \vert  \{ \emptyset \} \rangle}
  \def\ket2{\vert \vert \otimes \{ R \} \rangle}

\documentstyle[12pt]{ioplppt}
\begin{document}
\title{ Effect of short range order on
electronic and magnetic properties  of disordered Co based alloys}
\author{\bf Subhradip Ghosh$^{\dagger}$, Chhanda Basu Chaudhuri$^{\dagger}$, Biplab Sanyal$^{\ddag}$
 and Abhijit Mookerjee$^{\dagger}$}
\address{$^{\dagger}$S.N.Bose National Centre for Basic Sciences, JD Block,
Sector 3, Salt Lake City, Calcutta 700091, India\\
$^{\ddag}$ Department of Physics, Brock University, St. Catharines, Ontario L2S 3A1, Canada.}

\begin{abstract}
We here study electronic structure and magnetic properties 
of disordered CoPd and CoPt alloys using
Augmented Space Recursion technique coupled with the tight-binding
linearized muffin tin orbital (TB-LMTO) method. Effect of short range ordering 
present in disordered phase of alloys
on electronic and magnetic properties has been discussed. 
We present results for magnetic moments, Curie temperatures 
and electronic band energies with varying degrees of short range order for
different concentrations of Co and try to understand
and compare the magnetic properties and ordering phenomena in these systems.
\end{abstract}
{\parindent 0pt
\baselineskip 25pt

\section{Introduction}

The magnetic and chemical interactions in solid solutions, their
interdependence and the role they play in determining the electronic
and magnetic properties of  transition metal alloys have been the subject of
extensive experimental investigation \cite{kn:cad}.
Several phenomenological models based on
 statistical thermodynamic aspects of phase stability are 
available to describe the interplay between magnetism and spatial
order \cite{kn:sato} -\cite{kn:hen}.

Apart from this, there is one more approach of understanding the interplay
between magnetism and ordering in transition metal alloys which involves
investigation of the influence of local environment on electronic and
magnetic properties of these alloys. A considerable amount of literature  
exists concerning the local(short-range)order in
transition metal alloys obtained through measurements of X-ray or
neutron diffuse scattering, nuclear magnetic resonance and Mossbauer
spectroscopies \cite{kn:mir,kn:cad2,kn:mir1,kn:cad3}. In order to explain the
experimental results and describe the inhomogeneous character of magnetism
in these alloys many phenomenological models \cite{kn:jac}
as well as electronic structure calculations based on both zero and finite
temperature models\cite{kn:has} have been elaborated. The effect of local environment in
disordered alloys can be described in terms of short-range order(SRO)
because the degree of SRO determines the extent to which spatial correlations
exist in these systems. This approach has been adopted by many workers in
recent times in the framework of {\it ab-initio} electronic structure calculations
\cite{kn:borici,kn:zun,kn:abri}.

The macroscopic state of SRO for a disordered binary alloy is characterized
by Warren-Cowley \cite{kn:cow} SRO parameter which 
is given by

\begin{equation}
\alpha_{r}^{AB} = 1-\frac{P_{r}^{AB}}{y} \nonumber
\end{equation}

where the $A$ atom is at the center of the $r^{th}$ shell, $y$ denotes the macroscopic 
concentration of species $B$ and $P_{r}^{AB}$ is the pair probability of
finding a $B$ atom anywhere in the $r^{th}$ shell around an $A$ atom.

In the above mentioned approach, the workers either calculated the SRO
parameters for a certain degree of disorder using first principles techniques
and investigated the effect on ordering behavior of the systems\cite{kn:zun,kn:staun} 
or extracted the SRO parameters from experiments
and observed its effect on electronic structure and properties\cite{kn:borici,kn:abri}.

In this communication, we present the effect of SRO on the magnetic properties
and the ordering behaviour of Co based alloys. For our investigations, we have
chosen Co$_{x}$Pt$_{1-x}$ and Co$_{x}$Pd$_{1-x}$ alloys. Both the systems
have been studied extensively over the years. In recent times they have
received special attention due to their potentiality of being used as a
recording medium in a new generation of storage devices. For these reasons
lots of work on optical and magneto-optical characterization of these systems
are available in recent literatures \cite{kn:uba}.  
Theoretical calculations include anisotropic electrical resistivity studies by Ebert {\it et al}
\cite{kn:ebert}, investigation of electronic structure and magnetic properties
of ordered CoPt alloys by Kashyap {\it et al} \cite{kn:kashyap}, study of
magnetism in disordered CoPt alloys by Ebert{\it et al}\cite{kn:ebert1} and
calculation of ground state properties of CoPt by Shick {\it et al}\cite{kn:shick}. 
But, the interesting problem of interrelations of magnetism and local
ordering has failed to draw much attention. The interplay between these two
phenomena is quite complicated which has been studied experimentally by
Sanchez {\it et al}\cite{kn:sanchez}. To our knowledge no such investigation
has been done so far for CoPd. Hence,in this work we make an attempt to
understand the influence of short-range order on magnetic and electronic
properties in these systems from a first principles viewpoint. Our purpose is
to understand and compare these iso-electronic systems with respect to their responses to
degree of short range ordering. In particular,
we look at the behaviour of partial and average magnetic moments, Curie
temperatures and band energies with varying alloy compositions and degrees of SRO.

\section{Theoretical Details}

Our calculations are based on the generalized augmented space recursion (ASR) technique \cite{kn:as,kn:saha,kn:bs}. 
The effective one electron Hamiltonian within the local spin density approximation (LSDA is constructed 
in the framework of the tight-binding linearized muffin tin orbitals (TB-LMTO) method \cite{kn:oka}.
The Hamiltonian is sparse and therefore suitable for the application of the recursion method of Haydock \etal \cite{kn:hhk}.
The ASR allows us to calculate the configuration of the Green functions including short ranged ordering in
the Hamiltonian parameters. It does so by augmenting the Hilbert space spanned by the tight-binding basis
by the configuration space of the random Hamiltonian parameters. The configuration average is expressed {\sl exactly}
as a matrix element in the augmented space.  Details of  this methodology has been presented in an earlier paper
\cite{kn:ppb}. Here we shall quote the key results of generalized TBLMTO-ASR for short-ranged ordering. 
The augmented space Hamiltonian with short ranged order is written as 

\begin{eqnarray}
\^H = H_{1} + H_{2}\sum_{R} P_{R} \otimes P_{\downarrow}^{R} +
H_{3}\sum_{R} P_{R} \otimes (T_{\downarrow\uparrow}^{R}+T_{\uparrow\downarrow}^{R}) \nonumber\\
 +H_{4}\sum_{R}\sum_{R'}T_{RR'}\otimes\unit+ \alpha H_{2}\sum_{R''} P_{R''}\otimes P_{\downarrow}^{1} 
\otimes (P_{\uparrow}^{R''}-P_{\downarrow}^{R''}) \nonumber\\
 +H_{5}\sum_{R''} P_{R''}\otimes P_{\downarrow}^{1}\otimes (T_{\uparrow\downarrow}^{R''}+T_{\downarrow\uparrow}^{R''}) \nonumber\\
 +H_{6}\sum_{R''} P_{R''}\otimes P_{\downarrow}^{1}\otimes (T_{\uparrow\downarrow}^{R''}+T_{\downarrow\uparrow}^{R''}) \nonumber\\
 +\alpha H_{2} \sum_{R''} P_{R''}\otimes(T_{\uparrow\downarrow}^{1}+T_{\downarrow\uparrow}^{1})\otimes
(P_{\uparrow}^{R''}-P_{\downarrow}^{R''}) \nonumber\\
 + H_{7}\sum_{R''} P_{R''} \otimes (T_{\uparrow\downarrow}^{1}+T_{\downarrow\uparrow}^{1})\otimes 
(T_{\uparrow\downarrow}^{2}+T_{\downarrow\uparrow}^{2})
\end{eqnarray}

where $R''$ belong to the set of nearest neighbours of the site labelled 1i, at which the local density
of states will be calculated. $P$'s and $T$'s are the projection and transfer operators either in the
space  spanned by the tight-binding basis $\{\vert R\rangle\}$ or the configuration space associated with the sites
R , $\{\vert \uparrow_R\rangle, \vert\downarrow_R\rangle \}$ as described in
\cite{kn:ppb}. The different terms of the Hamiltonian are given below.

\begin{eqnarray}
H_{1} = A(C/\Delta)\Delta_{\lambda} -(EA(1/\Delta)\Delta_{\lambda}-1) \nonumber\\
H_{2} = B(C/\Delta)\Delta_{\lambda} -EB(1/\Delta)\Delta_{\lambda} \nonumber\\
H_{3} = F(C/\Delta)\Delta_{\lambda} -EF(1/\Delta)\Delta_{\lambda} \nonumber\\
H_{4} = (\Delta_{\lambda})^{-1/2} S_{RR'}(\Delta_{\lambda})^{-1/2} \nonumber\\
H_{5} = F(C/\Delta)\Delta_{\lambda} [\sqrt{(1-\alpha)x(x+\alpha y)} + \sqrt{(1-\alpha)y(y+\alpha x)} -1] \nonumber\\
H_{6} = F(C/\Delta)\Delta_{\lambda} [y\sqrt{(1-\alpha)(x+\alpha y)/x} + x\sqrt{(1-\alpha)(y+\alpha x)/y} -1] \nonumber\\
H_{7} = F(C\Delta)\Delta_{\lambda} [\sqrt{(1-\alpha)y(x+\alpha y)} - \sqrt{(1-\alpha)x(y+\alpha x)} \nonumber\\
A(Z)=xZ_{A}+yZ_{B} \nonumber\\
B(Z)=(y-x)(Z_{A}-Z_{B}) \nonumber\\
F(Z)=\sqrt{xy}(Z_{A}-Z_{B})
\end{eqnarray}

$\alpha$ is the nearest neighbour Warren-Cowley parameter described earlier. $\lambda$ labels  the constituents.
$C$'s and $\Delta$'s are the potential parameters describing the atomic scattering properties of the constituents
 and $S$ is the screened structure constant describing the
underlying lattice which is fcc in the present case. For convenience, all the angular momentum labels have been suppressed, with
the understanding that all potential parameters are 9$\ \times\ $9 matrices. 
First of all, we note that in absence of short-ranged order ($\alpha$ = 0), the terms 
H$_{5}$ to H$_{7}$ disappear and the Hamiltonian reduces to the standard one described earlier \cite{kn:ppb}.

The initial TB-LMTO potential parameters are obtained from suitable guess potentials
as described in the article by Andersen {\it \etal} \cite{kn:ajs}. In subsequent iterations
the potentials parameters are obtained from the solution of the Kohn-Sham equation

\begin{equation}
\left\{ -\frac{\hbar^{2}}{2m} \nabla^{2} + V^{\nu\sigma} - E\right\} \phi^{\nu}_{\sigma}(r_{R}, E)
\; =\; 0 \end{equation}

where,

\begin{equation}
V^{\lambda\sigma}(r_{R})\;  = \; V_{core}^{\lambda\sigma}(r_{R}) + V_{har}^{\lambda\sigma}(r_{R})
                           + V_{xc}^{\lambda\sigma}(r_{R}) + V_{mad}
\end{equation}

\noindent here $\lambda$ refers to the species of atom sitting at $R$ and $\sigma$ the spin component.
The electronic position within the atomic sphere centered at $R$ is given by $r_{R}$ =$r-R$.
The core potentials are obtained from atomic calculations and are available for
most atoms.
For the treatment of the Madelung potential, we follow the procedure suggested by Drchal {\it \etal} \cite{kn:dkw}. 
 We choose the atomic sphere
radii of the components in such a way that they preserve the total volume on the
average and the individual atomic spheres are almost charge neutral. This ensures
that total charge is conserved,  but each atomic sphere carries no excess
charge. In doing so, we had to be careful so that the spheres do not overlap much to violate the
atomic sphere approximation. 

The local charge densities are given by :

\begin{equation}
\rho^{\lambda}_{\sigma}(r) \eq (-1/\pi) \Im m \sum_{L} \int_{-\infty}^{E_{F}} dE \ll G_{LL}^{\lambda,\sigma}(r,r,E)\gg
\end{equation}

Here $\lambda$ is either $A$ or $B$. The local magnetic moment is

\[ m^{\lambda} \eq \int_{r<R_{WS}} d^{3}r\; \left[\rho_{\uparrow}(r)\mns \rho_{\downarrow}(r)\right] \]

The Curie temperature T$_{C}$ can be calculated using Mohn-Wolfarth (MW) model \cite{kn:mw} from the expression 
$$  \frac{T_{C}^2}{T_{C}^{S^2}}+\frac{T_{C}}{T_{SF}} \mns 1 \eq 0$$
where,
T$_{C}^S$ is the Stoner Curie temperature calculated from the relation
\begin{equation}
\langle I(E_{F}) \rangle \int_{-\infty}^{\infty} N(E) \left(\frac{\delta f}{\delta E} \right) dE \eq 1
\end{equation}
$\langle $I(E$_{F}$)$\rangle$ is the concentration averaged Stoner parameter. The parameters of pure elements are 
 obtained from the earlier calculations
\cite{kn:jan}
, $N(E)$ is the density of states per atom per spin \cite{kn:gun}
and $f$ is the Fermi distribution function.
$T_{SF}$ is the spin fluctuation temperature given by,
\begin{equation}
T_{SF} \eq \frac{m^2}{10k_{B} \langle \chi_{0}\rangle} 
\end{equation}
$ \langle \chi_{0} \rangle$ is the concentration weighted exchange enhanced spin susceptibility at equilibrium and $m$ is
the averaged magnetic moment per atom.
$\chi_{0}$ (pure elements) is calculated using the relation by Mohn \cite{kn:mw} and Gersdorf
\cite{kn:ger}:

\[\chi_{0}^{-1} \eq  \frac{1}{2\mu_{B}^2}\left(\frac{1}{2N^\uparrow(E_{F})}\pls
\frac{1}{2N^\downarrow(E_{F})} \mns I\right)\]

$I$ is the Stoner parameter for pure elements and $N^\uparrow(E_{F})$ and $N^\downarrow(E_{F})$ are
the spin-up and spin-down partial density of states per atom at the Fermi level for each species in the alloy.

\section{Computational Details}

For all the calculations, we have used a real space cluster of 400 atoms and
an augmented space shell upto the sixth nearest neighbour from the starting
state. Eight pairs of recursion coefficients were determined exactly and the
continued fraction was appended with the analytic terminator of Luchini
and Nex \cite{kn:ln}. In an earlier paper, Ghosh {\it \etal} \cite{kn:gdm} have shown the
convergence of the related integrated quantities, like the Fermi energy, 
the band energy,  the magnetic moments and the charge densities,  
within the augmented space
recursion. The convergence tests suggested by the authors were carried
out to prescribed accuracies. We noted that at least eight pairs
of recursion coefficients were necessary to provide Fermi energies
and magnetic moments to required accuracies. We have reduced the 
computational burden of the
recursion in the full augmented space by using the local symmetries of the
augmented space to reduce the effective rank of the invariant subspace
in which the recursion is confined \cite{kn:sdm} and using the seed
recursion methodology \cite{kn:gm} with fifteen energy seed points
uniformly across the spectrum.
The exchange-correlation potential of Von Barth and Hedin has been used.
s, p and d orbitals were used to construct the basis functions and scalar
relativistic corrections were included. 

\section{Results and Discussion}

We have performed total energy calculations for CoPd and CoPt alloys for several concentrations of Co to obtain the ground
state lattice parameters. Energy convergence was set for 0.01 mRyd.
The results are shown in Fig. 1. It is seen that for both CoPd and CoPt, there is a deviation from Vegard's law values(shown
by dashed lines) though the trends are same. In both the cases, equilibrium lattice parameters decrease with the increase in Co concentrations.
Shick \etal \cite{kn:shick} obtained the equilibrium lattice parameter for Co$_{50}$Pt$_{50}$ to be 7.049 a.u. using fully relativistic TBLMTO-CPA
in frozen core approximation. Both their value and our calculated value 
of 6.921 a.u. are less than the experimental lattice parameter of 7.10 a.u. \cite{kn:expt}.
This  is  not surprising as LSDA invariably overestimates bonding.

Results for magnetic moments of Co$_{x}$Pd$_{1-x}$ are shown in Fig. 2(a)
while that of Co$_{x}$Pt$_{1-x}$ are shown in Fig. 2(b). 
It is seen that both Pd and Pt sites also acquire
some induced moments from Co. Local magnetic moments of Co go down with increasing Co concentration but the
changes are not significant.
This is observed in ordered alloys too \cite{kn:kashyap}. This is a signature of weak local environmental effects on electronic structure.
According to the calculation of Shick \etal, the averaged and partial moments of Co and Pt in Co$_{50}$Pt$_{50}$ are 1.066, 1.787 and 0.345 $\mu_{B}$
respectively. We get the values of 1.049, 1.852 and 0.24 $\mu_{B}$ for the same while both of the values for averaged magnetic moments are close to
the experimental value of 1.05 $\mu_{B}$ \cite{kn:expt}. Theoretical
results using other techniques are not available for Co$_{x}$Pd$_{1-x}$ systems. But,our results for
both the systems agree well with experimental  \cite{kn:expt}.
As expected the LSDA estimate of the exchange field and hence the local
magnetic moment is always larger than experimental values.

The value of Warren-Cowley SRO parameter for A$_{x}$B$_{y}$
alloy is given by --$(x/y) \le \alpha <$ 1 where $\alpha$=--$(x/y)$
implies full short-range ordering and $\alpha$=1 implies complete segregation.
In our case we have taken our $\alpha$= --0.2 which is valid for the whole
range of concentrations. The results for magnetic moments of CoPd and CoPt 
systems have been shown in dashed lines of Figs. 2(a) and 2(b) respectively.
The results show that the effect of SRO included through the given value of
$\alpha$ on average magnetic moment is pretty weak. The difference in
values of partial moments in the SRO state and fully disordered state is not
uniform across the concentration axis for both the systems though the average
moment in the SRO state is always less than that of fully disordered state. In
case of CoPd, there is a crossover of partial moment value of both Co and Pd
at certain concentrations with respect to the disordered value. At around 35$\%$, 
the Co partial moment in the SRO state becomes less than that of disordered phase and this
trend follows for the higher concentrations of Co. For Pd, however, the change is
observed at around 55$\%$ but the quantitative difference with disordered phase
in case of Pd is almost negligible. Exactly the same trend is observed in
case of CoPt systems.

Results for MW and Stoner Curie temperatures for CoPd and CoPt are shown
in Figs. 3(a) and 3(b) respectively. MW Curie temperatures for both
the systems are in good agreement with the experiments \cite{kn:expt}.
On the other hand Stoner Curie temperatures are highly overestimated. This
is not surprising since we should realize that Stoner Curie temperature 
measures the temperature at which the paramagnetic state becomes unstable
rather than the magnetic transition temperature. This overestimation is much
reduced in the MW model \cite{kn:mw} which combines two extreme theories-
the single particle excitation and collective particle excitations. Again,
the theoretical Curie temperatures are higher than experimental values
due to the same reason as described in case of magnetic moments.

The results for Curie temperatures in SRO state of CoPd and CoPt are shown
in Figs. 3(a) and 3(b) by dashed lines. The Stoner Curie temperatures for both
the systems are almost unaffected by SRO. The influence in MW Curie
temperatures is also less. Yet, there is a difference in behaviour 
(quantitatively) with respect to fully disordered state at around 50$\%$ for
both the systems. Around 50$\%$ of Co, the difference in magnitude of Curie
temperature of SRO state and disordered state changes from positive to 
negative value.

Figure 4 shows the partial densities of states for equi-atomic CoPd and CoPt alloys with
SRO parameter -1.0, 0.0 and 1.0. While going from the short-ranged ordering side (-1.0) to the segregation
side (1.0) we find distinct changes in local DOS. The DOS for majority and minority electrons
shift relative to each other and bring change in magnetic moments. For CoPd alloy, the average magnetic moment
is increased from 0.96 $\mu_{B}$/atom to 1.24 $\mu_{B}$/atom while going from $\alpha$=-1.0 to 1.0. The change is
from 0.88 $\mu_{B}$/atom to 1.09 $\mu_{B}$/atom in case of CoPt.
  
To have a complete understanding of the ordering tendency and effect of local
ordering on magnetism in these systems we now carry out calculations for the
full range of $\alpha$ at different concentrations. 
Figs. 5(a), 5(b) and 5(c) show the
panels containing results for magnetic moments, electronic band energies and MW Curie temperatures
for Co$_{20}$Pd$_{80}$, Co$_{50}$Pd$_{50}$ and Co$_{50}$Pd$_{50}$ respectively. 
It is observed
that while at 20$\%$ concentration of Co, Co partial moment decreases towards
the segregation side,it shows a reverse tendency at 50$\%$ and 80$\%$. The Pd
partial moment shows a rise towards the segregation side at 20$\%$ while at 50$\%$
and 80$\%$ it remains almost at a constant magnitude. To understand this
behavioral difference of Co moment at different concentrations we present
results of magnetic moment at 10$\%$ and 40$\%$ of Co in Figs. 6(a) and 6(b) respectively. 
The results for 10$\%$ mimic that of 20$\%$ but the 40$\%$ case almost follows
the higher concentration trends. This can be understood in the following way:
As the system goes from ordering to the segregation side, more and more Co atoms
club together to build up magnetic moment of Co but at lower concentrations
($<$40$\%$) a Co atom finds itself in a completely non-magnetic Pd surrounding.
Therefore the situation is like a magnetic impurity in a non-magnetic host
which instead of building up rather subdues its moment as it goes towards
the segregation side. 

The middle panels containing the results for the band energies show that
at 20$\%$ the system shows a tendency towards segregation while at 50$\%$ and
80$\%$ the tendency is towards ordering. To locate the region of the transition,
figure 6(b) can be investigated which presents results on band energy for
40$\%$ Co. It is seen that at this concentration the system shows tendency
towards segregation which means that the ordering behavior of the system
changes between 40$\%$ and 50$\%$ of Co concentration.

The bottom panels show the variation of Curie temperature with SRO
parameter. At 20$\%$ and 50$\%$ concentrations
Curie temperatures are higher towards ordering side while the trend is
opposite at 80$\%$.In other words, at 20$\%$ and 50$\%$ ferromagnetic phases
are stable upto higher temperatures in the ordering side while at 80$\%$ 
they are stable upto higher temperatures in the segregation side.

Figs. 7(a), 7(b) and 7(c) present the results for the same properties
but for CoPt alloys.
The nature of variation of the moments are exactly same as those of CoPd and
hence can be explained using the same logic. 
The results for the band energies show that at 20$\%$ and
50$\%$ of Co, the system shows the tendency towards ordering while at 80$\%$ it tends
to segregate. Once again to locate the region of transition the bottom panels of Figs. 8(b) and
8(c) are referenced. Here, the change in ordering behavior is observed 
at 60$\%$ of Co which indicates that unlike CoPd, this system has a tendency to 
segregate between 50$\%$ and 60$\%$ of Co.
From the bottom panels of Fig. 7(a-c), it is seen that at all concentrations, Curie temperature is
higher at higher band energy sides. It is indicative of the possibility that
the ferromagnetic phases are stable upto a lower temperature at the minimum
energy state of this system.

\section{Conclusions}

We have studied the effects of short range order on the
magnetic and electronic properties of the Co$_{x}$Pd$_{1-x}$ and
Co$_{x}$Pt$_{1-x}$ alloys using fully self consistent first principles techniques.
Our results for completely disordered phases agree reasonably well with
the experiments. The effect of SRO on magnetic moments, electronic band energies
and Curie temperatures have been investigated in detail. CoPt shows a
tendency to go to ordering state from clustering(segregation) state at around
60$\%$ of Co while CoPd shows this tendency at around 40$\%$. The response of
Curie temperature to short-range ordering is linear in CoPt in the sense that
at all concentrations it attains higher value at the energetically higher
SRO states. For CoPd, the response is not that linear. At 20$\%$ and 80$\%$
concentrations higher values are observed at energetically higher SRO states
while at 50$\%$ higher values are observed at energetically lower SRO states.

\section*{Acknowledgments}
CBC would like to thank the CSIR, India for financial assistance.

\section*{Figure Captions}
\begin{description}
\item[Figure 1] Equilibrium lattice parameters (in a.u.) vs. concentration of Co for (top) CoPd
(bottom) CoPt alloys. The circles represent the calculated values whereas the dashed lines are for
Vegard's law values.
\item[Figure 2] Partial and averaged magnetic moments (in Bohr-magnetons/atom) vs. concentration of Co 
in (a) CoPd (b) CoPt alloys. The full line is for disordered case and the dotted one for
SRO state with $\alpha$=-0.2.
The symbols represent : filled squares, filled circles and filled diamonds  
are for Co partial moments, averaged moments and Pd partial moments respectively. Diamonds represent
the experimental values of average magnetic moment in fully disordered case.
\item[Figure 3] Curie temperature (in Kelvin) vs. concentration of Co in (a) CoPd and (b) CoPt alloys.
Panel(a): MW Curie temperature results. 
Full line represents fully disordered case. Dashed line represents SRO state
characterized by $\alpha$=-0.2. Diamonds represent experimental points for fully disordered
case. 
Panel(b): Stoner Curie temperature results. Full and dashed lines refer to the same results
as in (a). 
\item[Figure 4] Spin and component projected local densities of states/atom of (a)-(c) Co$_{50}$Pd$_{50}$ and (d)-(f)
Co$_{50}$Pt$_{50}$ alloys for SRO parameter equal to [(a) and (d)] -1.0 [(b) and (e)] 0.0 [(c) and (f)] 1.0. 
In all cases, the solid lines are for Co and the dashed lines are for Pd/Pt components. Vertical lines
show the positions of Fermi levels.
\item[Figure 5] Variation of properties of CoPd alloys with Warren-Cowley short
ranged order parameter. (a) Co$_{20}$Pd$_{80}$ (b) Co$_{50}$Pd$_{50}$ and (c) Co$_{80}$Pd$_{20}$ alloys.
Panels : (top) variation of partial and average magnetic moments (in Bohr magnetons/atom); symbols : filled circles, triangles and squares
are for Co,average and Pd moments respectively.
(middle) variation of band energies
(in Ryd.) (bottom) variation of MW Curie temperatures (in Kelvin).
\item[Figure 6] Variation of properties with SRO parameter for (a) Co$_{10}$Pd$_{90}$ and (b) Co$_{40}$Pd$_{60}$ alloys.
Panels : (top) variation of partial and average magnetic moments ; symbols : filled circles, triangles and squares
are for Co, average and Pd moments respectively.
(bottom) variation of band energy. 
\item[Figure 7] exactly as described for fig. 5 but for CoPt alloys.
\item[Figure 8] Variation of properties with SRO parameter for (a) Co$_{10}$Pt$_{90}$ (b) Co$_{40}$Pt$_{60}$ 
and (c) Co$_{60}$Pt$_{40}$ alloys.
Panels : (top) variation of partial and average magnetic moments ; symbols : filled circles, triangles and squares
are for Co, average and Pt moments respectively.
(bottom) variation of band energy. 

\end{description}

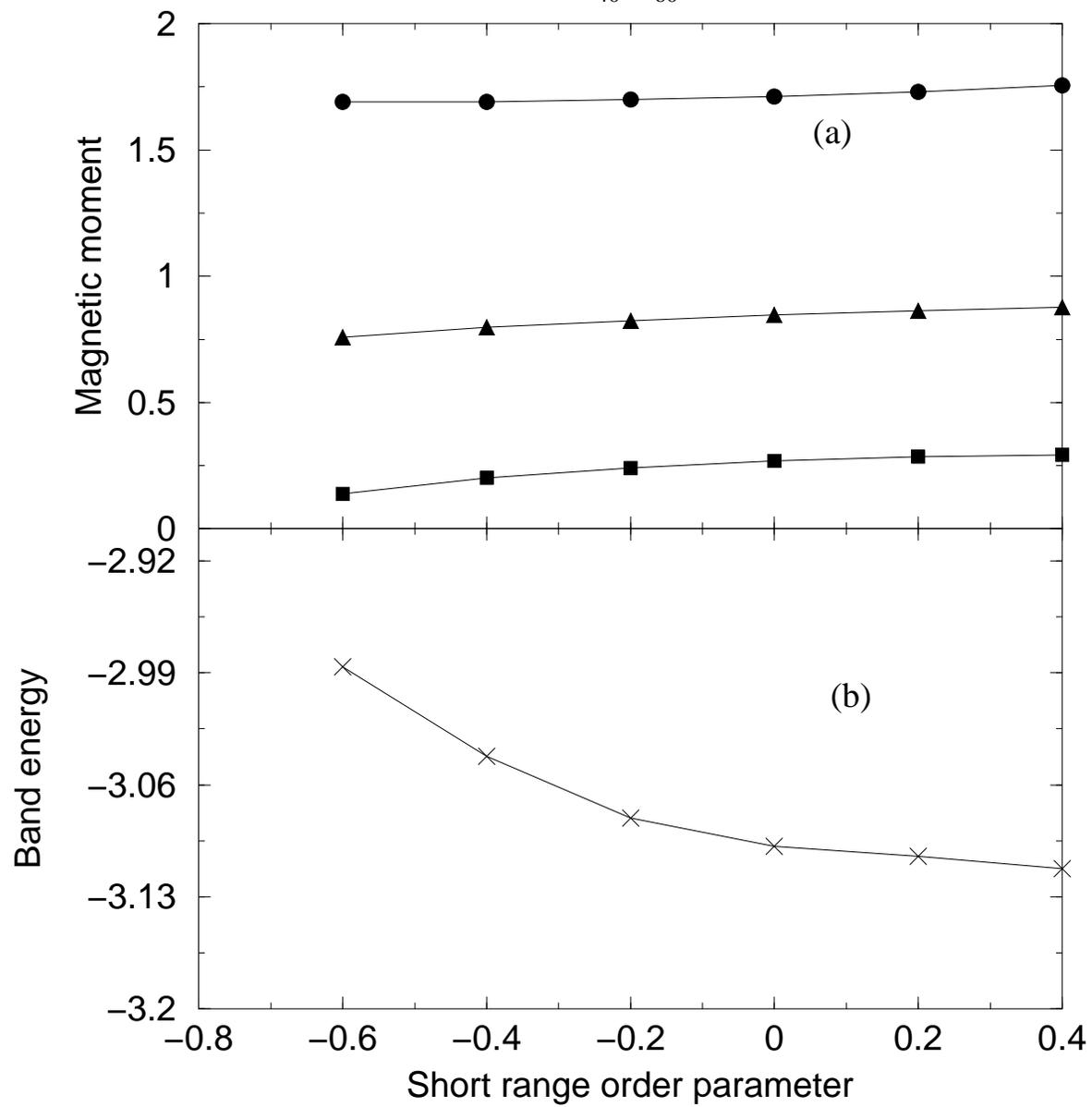
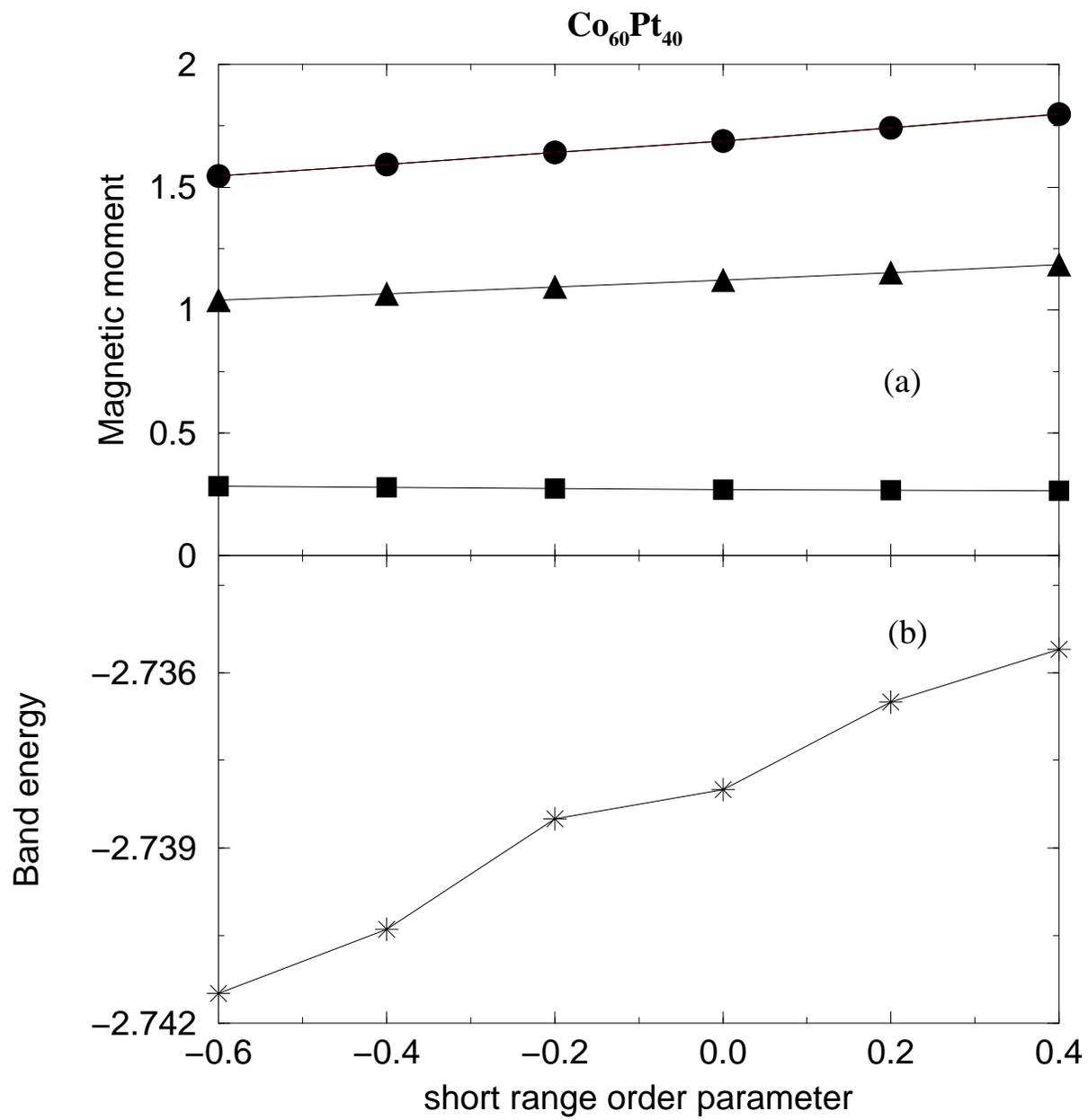
\end{document}